\documentclass[12pt,preprint]{aastex}

\newcommand{\feh} {\mbox{\rm [Fe/H]}}
\newcommand{\Mx} {\mbox{$M_{\rm X}$}}
\newcommand{\MV} {\mbox{$M_{\rm V}$}}
\newcommand{\MR} {\mbox{$M_{\rm R}$}}
\newcommand{\MI} {\mbox{$M_{\rm I}$}}
\newcommand{\Mu} {\mbox{$M_{\rm u}$}}
\newcommand{\Mg} {\mbox{$M_{\rm g}$}}
\newcommand{\Mr} {\mbox{$M_{\rm r}$}}
\newcommand{\Mi} {\mbox{$M_{\rm i}$}}
\newcommand{\Mz} {\mbox{$M_{\rm z}$}}

\newcommand{\gmr} {\mbox{$(g-r)_{\rm 0}$}}
\newcommand{\umg} {\mbox{$(u-g)_{\rm 0}$}}
\newcommand{\rmi} {\mbox{$(r-i)_{\rm 0}$}}
\newcommand{\imz} {\mbox{$(i-z)_{\rm 0}$}}

\begin{document}


\title{The Absolute Magnitudes of Red Horizontal Branch Stars in {\bf the} $ugriz$ System}
\author{Y.Q. Chen, G. Zhao and J.K. Zhao}

\altaffiltext{}{National Astronomical Observatories, Chinese
Academy of Sciences, Beijing, 100012, China; gzhao@bao.ac.cn.}

\begin{abstract}
Based on photometric data of the central parts of eight globular
clusters and one open cluster presented by An and his collaborators,
we select red horizontal branch (RHB) stars in the \gmr-$g_0$
diagram and make a statistical study of the distributions of their
colors and absolute magnitudes in the SDSS $ugriz$ system.
Meanwhile, absolute magnitudes in the Johnson $VRI$ system are
calculated through the translation formulae between $gri$ and $VRI$
in the literature. The calibrations of absolute magnitude as
functions of metallicity and age are established by linear
regressions of the data.

It is found that metallicity coefficients in these calibrations
decrease, while age coefficients increase,  from the blue $u$
filter to the red $z$ filter. The calibration of $\Mi=
0.06\feh+0.040t+0.03$ has the smallest scatter of 0.04 $mag$, and
thus $i$ is the best filter in the $ugriz$ system when RHB stars
are used for distance indicators. The comparison of the $\MI$
calibration from our data with that from red clump stars indicates
that the previous suggestion that the $I$ filter is better than
the $V$ filter in distance determination may not be true because
of its significant dependence on age.

\end{abstract}


\section{Introduction}
Stars with constant absolute magnitude are of high interest in
astrophysics because they can be used as distance indicators. For a
long time, horizontal branch stars have been widely adopted to
determine distances of globular clusters and nearby galaxies. At
present, various calibrations of absolute magnitude with metallicity
are established by many works. For the low metallicity range of
$\feh \leq -1.5$, absolute magnitudes in the $V$ band of RR Lyrae
stars in globular clusters and in the field are investigated by
using different methods and from different data. The results are not
always consistent as shown in the review paper by Sandage \& Tammann
(2006). The coefficient $a$ in the linear relation of $\MV =
a\feh+c$ varies from 0.18 (Carretta et al. 2000) to 0.37 (Feast
1997) and the constant $c$ varies from 0.74 to 1.13 mag. In the
solar neighborhood, absolute magnitudes in the $I$ band of red clump
(hereafter RC) stars in the metallicity range of $\feh \geq -0.5$
are well studied after the release of HIPPARCOS parallaxes (e.g.
Stanek \& Garnavich 1998; Zhao et al. 2001; Groenewegen 2008). Most
calibrations are consistent with a relation of $\MI =
0.13\feh-0.23$, and it is suggested that calibrations of absolute
magnitude in the $I$ band are better than those in the $V$ band
because they have a small dependence on metallicity. But, these
calibrations do not take into account age dependence, probably due
to the difficulty of deriving ages for RC stars. In view of this
situation, it is interesting to investigate if the calibrations
established from RR Lyrae stars with $\feh < -1.5$ are valid in the
mild-metallicity range of $-1.5 \leq \feh \leq -0.5$. Moreover, one
may ask how the calibrations of RR Lyrae and RC stars are connected
at a fixed metallicity in between.  In this respect, RHB stars may
be important for establishing the calibrations for the mild
metallicity population because they are significantly populated in
the color-magnitude diagrams (hereafter CMDs) of mild metal poor
globular clusters.

In this work, we aim to determine absolute magnitudes for RHB stars
based on the photometric data of 17 globular clusters and three open
clusters in the $ugriz$ system presented in An et al. (2008). The
main goals of this study are as follows. First, the calibrations of
absolute magnitude can be established in the $ugriz$ system and they
are useful in distance determinations of interesting populations
based on the large database of the SDSS photometric survey.
Moreover, wavelength coverage of $ugriz$ filters are narrower than
those of Johnson $V$ and $I$ filters ,and thus they provide more
accurate magnitudes of RHB stars. Second, it is possible to
investigate which filter, among the five $ugriz$ filters, is the
best distance indicator and how the absolute magnitude varies with
metallicity and age. Finally, it is interesting to compare these
calibrations and decide which calibration, in which filter, is the
best for a particular population. Should we use different
calibrations for a different metallicity range? Will the nonlinear
calibrations with metallicity improve the precision of absolute
magnitude determination? Is it necessary to include the age term in
the calibrations? How are these calibrations adopted in different
cases?

\section{The selection of RHB stars}
The photometric data of 17 Galactic globular clusters and three
open clusters in the SDSS $ugriz$ system are taken from An et al.
(2008) who applied the DAOPHOT/ALLFRAME programs to the central
parts of clusters. These programs can provide more reliable
$ugriz$ photometric data for crowded cluster fields than those of
the standard SDSS pipeline.

The selection of RHB stars in each cluster is based on the following
procedures. As described in An et al. (2008), some clusters are
observed more than one time and there are some overlapping regions
between different observation runs. In view of this, the first step
is to identify common stars in different observation runs and to
provide a clear sample so that the same star observed for more than
one time will not be considered as two or more stars. In the present
work, stars with the same positions, i.e. both $\Delta |ra\,
cos(dec)|$ and $\Delta dec$ being less than 0.5 arcsec, and the same
colors, i.e. $\Delta g$ being less than 0.2 mag, in different
observation runs are considered to be the same object and we include
only one of them in the final sample. Then, we select stars within
the tidal radius of the cluster and plot the CMD in different
filters based on the reddening taken from Harris (1996), 2003
February version, which is presented in Table 1 of An et al. (2008).
In the whole paper, the colors and magnitudes that we presented are
reddening-corrected parameters.

Among the 20 clusters, eight globular clusters and one open cluster
(presented in Table 1) show clear clumping of RHB stars. The nine
clusters cover the metallicity range of $-1.7 < \feh < +0.4$ and the
age range of 8-12 Gyr. The ages of the clusters are taken from
Salaris \& Weiss (2002). As shown in Figure 1, the ages of five
clusters increase with decreasing metallicity (they are classified
as group 1) while four clusters (Pal 3,4,5,14) are significantly
younger for their metallicity at $\feh \sim -1.5$ (they are
classified as group 2). It is emphasized that clusters in group 2
are important, which makes it possible to investigate the age effect
on the absolute magnitude in the present work. The two groups of
clusters will be considered separately in the following analysis
when necessary.

\begin{figure}
\epsscale{1.0} \plotone{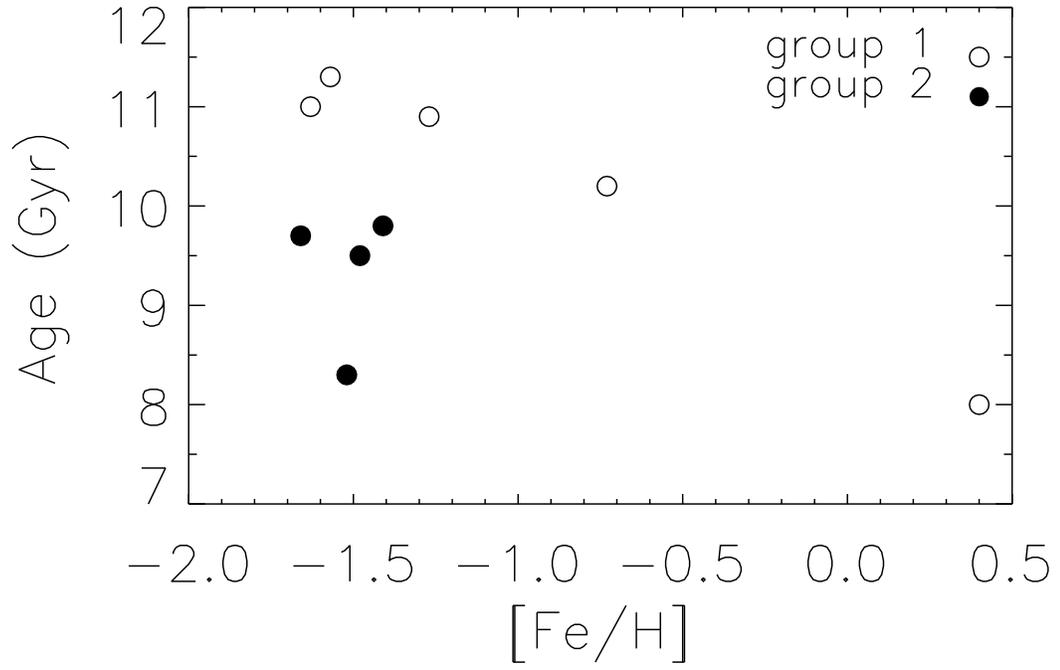} \caption{Age-metallicity diagram for
nine clusters in group 1 (open) and group 2 (filled).} \label{fig1}
\end{figure}

Usually, RHB stars in the CMD of each cluster, shown in Figure 2,
are well separated from BHB stars, due to the presence of a gap in
between, and from RGB stars, due to their low metallicity and/or old
age. But we should keep in mind that the so-called RHB stars in this
paper may include some RR Lyrae stars and even redder RC stars in
clusters because we do not have suitable data or criteria to exclude
them. Fortunately, it is expected that they will not affect the
result of this paper in regards to statistics due to the small
number and similar magnitudes in the HB of clusters.

The distributions of RHB stars are slightly different in the CMDs
based on different colors and filters in the $ugriz$ system. As
shown in Figure 2, the \gmr-$g_0$ diagram of M5 is the best diagram
to pick out RHB stars (within the defined box) in the sense that the
gap between the BHB and RHB populations is clear and the RHB
population is quite flat as \gmr\ varies. The RHB population is also
clear in the \gmr-$r_0$ diagram. However, it is difficult to pick
out RHB stars based on the \umg-$u_0$ and \umg-$g_0$ diagrams
because \umg\ is a significantly metallicity-sensitive index, and
RHB and BHB populations have the overlapping \umg\ colors at 1.0-1.2
mag. Also, it is not easy to define the RHB population from the
\rmi-$i_0$ and the \imz-$z_0$ diagrams because the gaps between RHB
and BHB/RGB are not so clear. Moreover, the photometric precisions
of $g_0$ and $r_0$ filters are generally higher than those of $u_0$,
$i_0$ and $z_0$ filters in the SDSS survey. Therefore, the
\gmr-$g_0$ diagram has the advantage of picking out RHB stars from
these clusters.

In order to define the colors and magnitude ranges of RHB stars in
the \gmr-$g_0$ diagram, we adopt a critical radius ($Rs$ in Table 1)
for each cluster, within which stars are selected to be plotted in
the CMD. The exception is Pal 4, for which we select all stars
presented in the An et al. (2008) paper without using any critical
radius because there are not enough stars even within the tidal
radius. The half diameter from van den Bergh (2006) is adopted to be
the critical radius for open cluster NGC\,6791. For the remaining
clusters, a critical radius between half light radius and tidal
radius is chosen so that the RHB population in the CMD will become
more clear. Then, we can define colors and magnitude ranges of RHB
stars for clusters by eye. As we show later, the resulting absolute
magnitudes will not be affected by the exact edges of the defined
colors and magnitudes of RHB populations. Figure 3 shows the defined
boxes of RHB populations based on the \gmr-$r_0$ diagrams for the
other eight clusters except for M5 which is already shown in Figure
2.

\begin{figure*}
\epsscale{1.0} \plotone{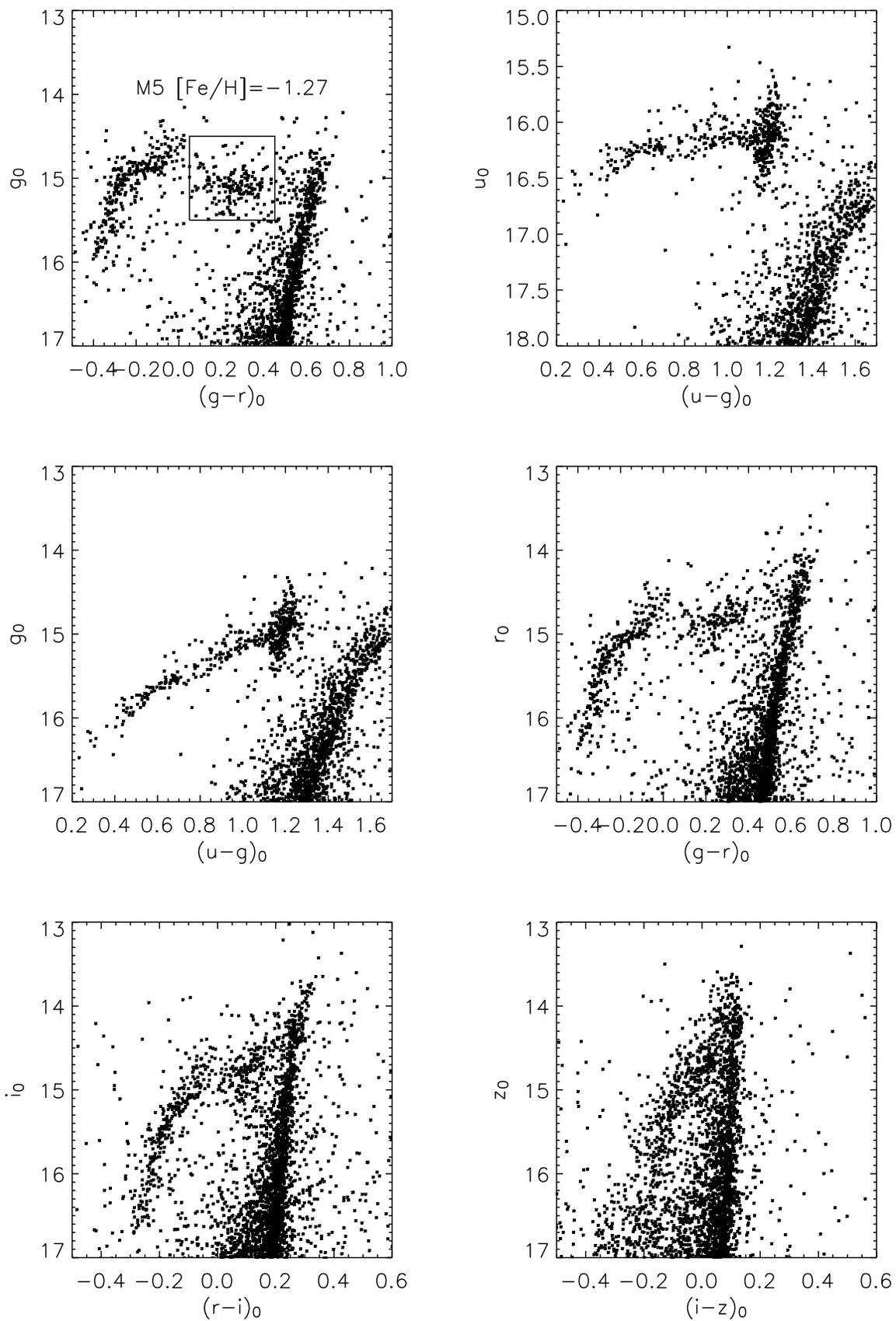} \caption{CMDs of M5 based on
different filters where stars within the tidal radius are plotted.}
\label{fig2}
\end{figure*}

\begin{figure*}
\epsscale{0.6} \plotone{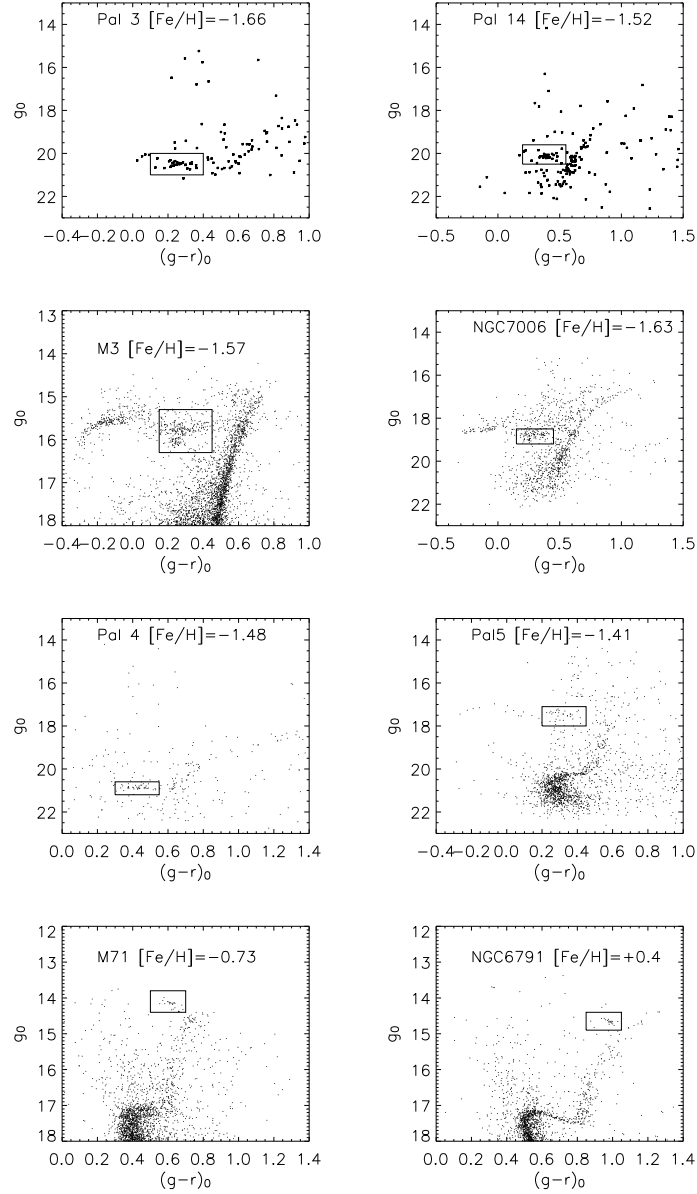} \caption{$\gmr$-$g_0$ diagrams of
the eight clusters.} \label{fig3}
\end{figure*}

\begin{table}
\caption{Critical radii ($Rs$, arcmin), \gmr\ color ranges and
$g_0$ magnitude ranges of RHB stars, together with metallicities,
ages, reddening and distance modulus of nine clusters.}
\label{tb:sel} \setlength{\tabcolsep}{0.1cm}
\begin{tabular}{rrrrrrrrrr}
\noalign{\smallskip}
\hline
Cluster  & $Rs$ & \multicolumn{2}{c}{\gmr range}  & \multicolumn{2}{c}{$g_0$ range}  & $\feh$ & Age &E(B-V) & $(m-M)_0$ \\
\noalign{\smallskip}
  Pal3    &   4.56 & \multicolumn{2}{c}{0.10 - 0.40}& \multicolumn{2}{c}{ 20.0 -  21.0} & -1.66 &  9.7 & 0.04 & 19.84\\
  NGC7006 &   4.23 & \multicolumn{2}{c}{0.15 - 0.45}& \multicolumn{2}{c}{ 18.5 -  19.2} & -1.63 & 11.0 & 0.05 & 18.09\\
  M3      &   9.83 & \multicolumn{2}{c}{0.15 - 0.45}& \multicolumn{2}{c}{ 15.3 -  16.3} & -1.57 & 11.3 & 0.01 & 15.09\\
  pal14   &   2.56 & \multicolumn{2}{c}{0.20 - 0.55}& \multicolumn{2}{c}{ 19.6 -  20.5} & -1.52 &  8.3 & 0.04 & 19.35\\
  pal4    &   3.33 & \multicolumn{2}{c}{0.30 - 0.55}& \multicolumn{2}{c}{ 20.6 -  21.2} & -1.48 &  9.5 & 0.01 & 20.19\\
  pal5    &  16.28 & \multicolumn{2}{c}{0.20 - 0.45}& \multicolumn{2}{c}{ 17.1 -  18.0} & -1.41 &  9.8 & 0.03 & 16.83\\
  M5      &  28.40 & \multicolumn{2}{c}{0.05 - 0.45}& \multicolumn{2}{c}{ 14.5 -  15.5} & -1.27 & 10.9 & 0.03 & 14.37\\
  M71     &   4.24 & \multicolumn{2}{c}{0.50 - 0.70}& \multicolumn{2}{c}{ 13.8 -  14.4} & -0.73 & 10.2 & 0.25 & 13.02\\
  NGC6791 &   5.00 & \multicolumn{2}{c}{0.85 - 1.05}& \multicolumn{2}{c}{ 14.5 -  15.5} &  0.40 &  8.0 & 0.16 & 13.02\\
\hline \noalign{\smallskip}
\end{tabular}
\end{table}

\section{Results and Discussions}
\subsection{The color and absolute magnitudes of RHB stars}

With the defined colors and magnitude ranges of RHB stars in Figures
2 and 3, we select all possible RHB candidates. Then, Gaussian
fittings to the color and magnitude distributions of the selected
RHB stars provide the centering values and their scatters. As an
example, Figure 4 shows the Gaussian fits to the distributions of
\gmr, $g_0$, $u_0$, $r_0$, $i_0$, and $z_0$ for M5 at $\feh=-1.27$.
Usually, these distributions have sharp peaks and do not exactly
obey Gaussian functions. Thus, the scatters from these Gaussian fits
may be meaningless, but the centering values are generally
consistent with these sharp peaks. Meanwhile, it is clear that the
distributions of $z_0$ magnitudes are broader than those of $ugri$
magnitudes. We note that photometric errors in the SDSS survey
cannot explain this difference because the errors in $ugriz$ are
about 0.01-0.02 mag for RHB stars in M5 with $g \sim 15\, mag$,
which are significantly smaller than the broadening widths of
0.1-0.2 mag found in Figure 4.

\begin{figure}
\epsscale{1.0} \plotone{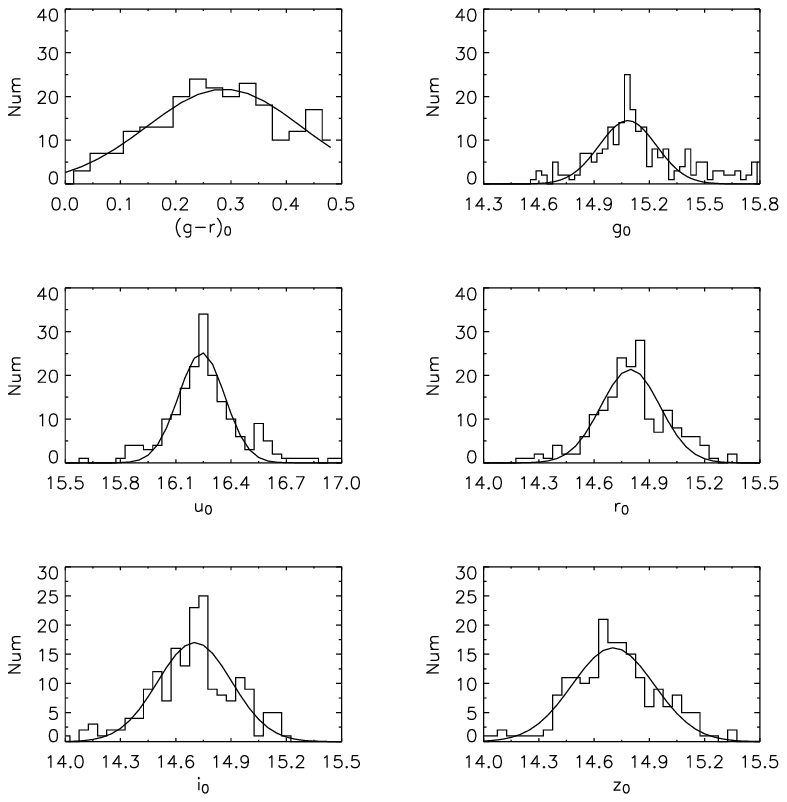} \caption{The \gmr\ color and
$ugriz$ magnitude distributions, together with their Gaussian
fittings, of the selected RHB stars for M5.} \label{fig4}
\end{figure}

With the distance modulus from Table 1 of An et al. (2008), $ugriz$
magnitudes (after reddening correction) are translated into absolute
magnitudes. Table 2 presented \gmr\ colors, absolute magnitudes and
their scatters estimated from the widths of the Gaussian
distributions for the clusters. It shows that $\Mu$ and $\Mg$ have
quite large ranges and the deviations reach 2.7 and 1.0 mag
respectively in the metallicity range of $-1.7 \leq \feh \leq +0.4$.
The variations of $\Mr$ and $\Mz$ reduce to 0.35-0.45 mag and $\Mi$
shows the smallest deviation of 0.16 mag in this metallicity range.
By using the relation of Chonis \& Gaskell (2008), it is possible to
transfer $\Mg$, $\Mr$ and $\Mi$ into $M_{V}$, $M_{R}$, and $M_{I}$,
and they are also presented in Table 2.

\begin{table*}
\caption{The centering colors and absolute magnitudes, together
with their scatters, of RHB stars in nine clusters. The
$\MV$,$\MR$ and $\MI$ are also presented.}
\label{tb:Mx} \setlength{\tabcolsep}{0.10cm}
\begin{tabular}{rrrrrrrrrrrrrrrr}
\noalign{\smallskip}
\hline
Cluster  & $\gmr$  & $\sigma$ & $\Mu$ & $\sigma \Mu$ & $\Mg$ & $\sigma \Mg$ & $\Mr$ & $\sigma \Mr$ & $\Mi$ & $\sigma \Mi$ & $\Mz$ & $\sigma \Mz$  & $\MV$ & $\MR$ & $\MI$\\
\noalign{\smallskip} \hline
  Pal3  &   0.25  & 0.08  &1.71  & 0.25  &0.60  & 0.13  &0.28  & 0.08  &0.28  & 0.08  &0.28  & 0.10 &  0.40 &    0.12 &  $-$0.09\\
  NGC7006  &0.31  & 0.13  &1.80  & 0.15  &0.60  & 0.30  &0.37  & 0.22  &0.27  & 0.26  &0.27  & 0.26 & 0.45  &   0.18 &  $-$0.13\\
  M3    &   0.28  & 0.12  &1.80  & 0.22  &0.64  & 0.21  &0.40  & 0.30  &0.31  & 0.30  &0.31  & 0.32 & 0.49  &   0.22 &  $-$0.09 \\
  pal14  &  0.40  & 0.12  &1.91  & 0.27  &0.77  & 0.06  &0.35  & 0.15  &0.15  & 0.16  &0.07  & 0.22 & 0.51  &   0.14 &  $-$0.28\\
  pal4  &   0.40  & 0.09  &1.88  & 0.35  &0.60  & 0.13  &0.20  & 0.08  &0.17  & 0.08 &$-$0.05& 0.15 & 0.36  &   0.04 &  $-$0.21\\
  pal5  &   0.29  & 0.08  &1.71  & 0.07  &0.62  & 0.24  &0.25  & 0.19  &0.25  & 0.19  &0.14  & 0.27 & 0.39  &   0.09 &  $-$0.12\\
  M5  &     0.27  & 0.15  &1.84  & 0.14  &0.67  & 0.17  &0.39  & 0.17  &0.30  & 0.20  &0.30  & 0.23 & 0.50  &   0.21 &  $-$0.10\\
  M71  &    0.60  & 0.02  &2.67  & 0.08  &1.09  & 0.04  &0.48  & 0.03  &0.23  & 0.04  &0.16  & 0.04 &  0.72 &    0.25&   $-$0.22\\
  NGC6791  &0.98  & 0.03  &4.42  & 0.27  &1.60  & 0.07  &0.65  & 0.06  &0.31  & 0.05  &0.15  & 0.07 & 1.03  &   0.40 &  $-$0.17\\
\noalign{\smallskip} \hline
\multicolumn{16}{c}{New reddening and distance modulus suggested by An et al. (2009)}\\
\noalign{\smallskip} \hline
  M71  &    0.62  & 0.09  &2.71  & 0.07  &1.20  & 0.15  &0.64  & 0.15  &0.36  & 0.11  &0.27  & 0.07 &  0.86 &    0.40&   $-$0.11\\
  NGC6791  &1.04  & 0.02  &4.65  & 0.26  &1.82  & 0.07  &0.80  & 0.06  &0.44  & 0.08  &0.24  & 0.06 & 1.21  &   0.54 &  $-$0.06\\
\hline
\noalign{\smallskip}
\end{tabular}
\end{table*}

\subsection{Absolute Magnitudes as Functions of metallicity and age}

Absolute magnitudes in $ugriz$ filters of RHB stars are investigated
as functions of the cluster's metallicity and age in Figure 5. It
shows that $\Mu$ has a good correlation with the cluster's
metallicity and has only weak dependence on the cluster's age. $\Mg$
and $\Mr$ also increase with the increasing metallicity while $\Mi$
and $\Mz$ do not show significant variation in the metallicity range
of $-1.7 < \feh < +0.4$.  For clusters with $\feh < -1.0$,  $\Mu$
and $\Mg$ decrease, while $\Mi$ and $\Mz$ increase with increasing
age, and $\Mr$ has little dependence on the cluster's age.

\begin{figure*}
\epsscale{1.0} \plotone{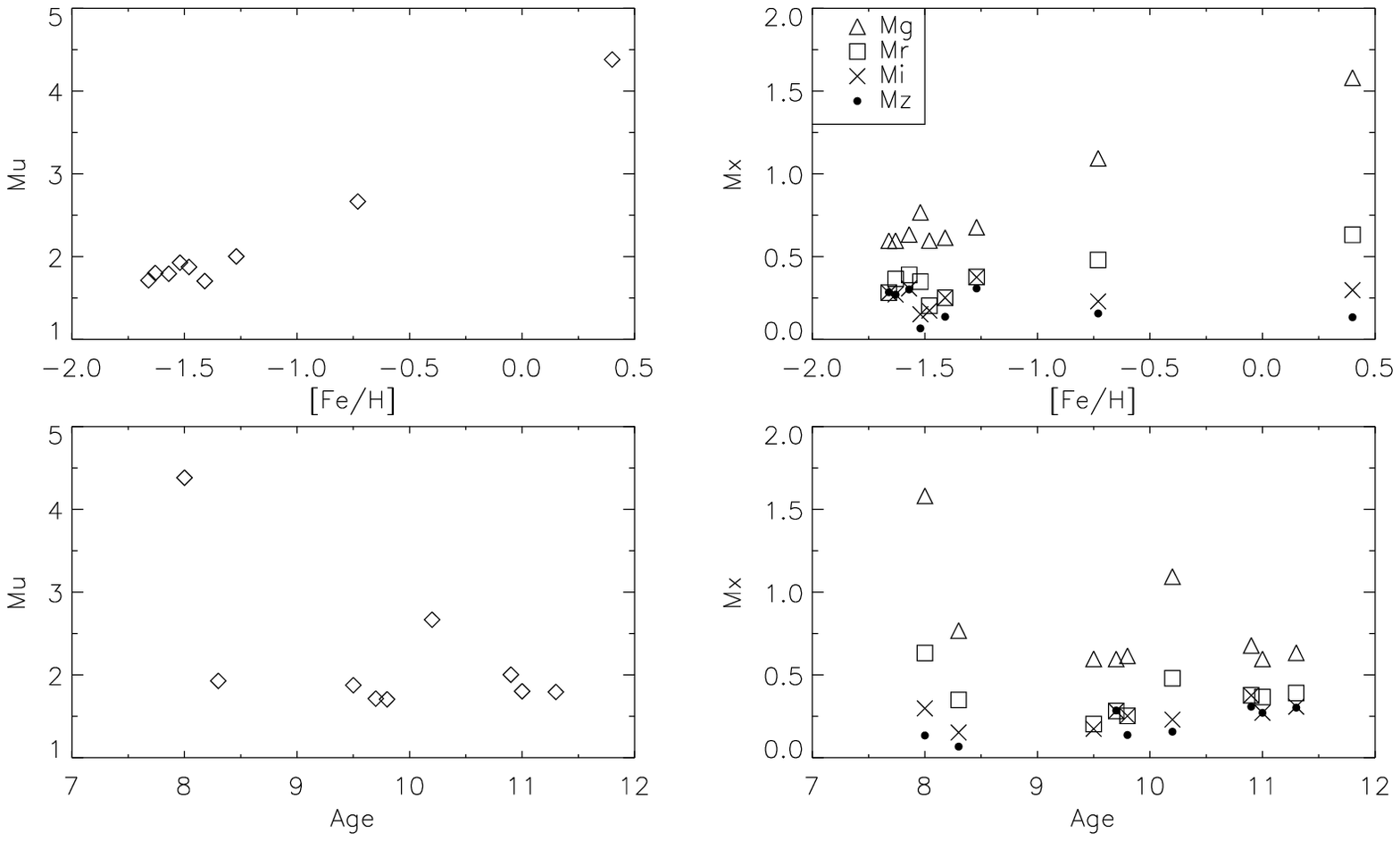} \caption{$\Mx$, ($x=u,g,r,i$,and $z$
indicated by diamonds, triangles, squares, crosses and circles
respectively) vs. $\feh$ and age.} \label{fig5}
\end{figure*}

In order to estimate this sensitivity quantitatively, we perform
multivariate fittings to the data in the formula  $Mx=a \feh + b t +
c$, where $t$ indicates the cluster's age and $x=u,g,r,i,z, V, R,$
and $I$. The coefficients and scatters of the calibrations are shown
in Table 3. Note that the metallicities of globular clusters in the
present work are taken from Harris (1996), Feb. 2003 version. They
are found to agree well with those from Kraft \& Ivans (2003). As
pointed out by An et al. (2009), the metalicities in  Kraft \& Ivans
(2003) are more reliable because they are derived from FeII lines,
which usually do not suffer from NLTE effects and they agree well
with those of Zinn \& West (1984). But, we note that the metallicity
for M71 in Zinn \& West (1984) is too high ($-$0.58 vs. $-$0.81) as
compared with that from Kraft \& Ivans (2003). The metallicity of
M71 in Harris (1996), 2003 February version, is in between. We have
checked that the results present in Table 3 will not be altered
significantly when the metallicities from either Zinn \& West (1984)
or Carretta \& Graton (1997) are used.

\begin{table}
\caption{The coefficients, $a$, $b$, $c$, and their errors as well
as the scatters of the calibrations in the formula of $M=a \feh +
b t + c$. The reddening and distance modulus are taken from Harris
(1996) 2003 version for globular clusters and from An et al.
(2008) for NGC6791.} \label{tb:sel} \setlength{\tabcolsep}{0.1cm}
\begin{tabular}{rrrrrrr}
\noalign{\smallskip}
\hline
Filters  & $a$ & $b$ & $c$ & $\sigma_a$ & $\sigma_b$ & $\sigma$\\
\noalign{\smallskip}
$\Mu$ &     1.243 &  $-$0.046 &   4.158 &   0.095 &   0.055 &   0.132\\
$\Mg$ &    0.469 &  $-$0.026 &   1.614 &   0.047 &   0.027 &   0.065\\
$\Mr$ &    0.192 &   0.030 &   0.311 &   0.043 &   0.025 &   0.060\\
$\Mi$ &    0.064 &   0.049 &  $-$0.148 &   0.034 &   0.020 &   0.047\\
$\Mz$ &    0.041 &   0.082 &  $-$0.576 &   0.065 &   0.037 &   0.090\\
$\MV$ &    0.306 &   0.007 &   0.838 &   0.043 &   0.025 &   0.060\\
$\MR$ &    0.157 &   0.035 &   0.025 &   0.033 &   0.019 &   0.046\\
$\MI$ &    0.021 &   0.056 &  $-$0.671 &   0.047 &   0.027 &   0.065\\
\hline \noalign{\smallskip}
\end{tabular}
\end{table}

When we adopt the reddening and distance modulus from Grundahl et
al. (2002) for M71, the absolute magnitudes deviate by 0.04-0.16
mag. For NGC6791, we adopt E(B-V)=0.16 and $(m-M)_0$=13.02, which
agree with the results from Carney et al. (2005: 0.14/13.07) based
on JHK photometry, Anthony-Twarog et al. (2007: 0.16/13.10) based on
uvby photometry and Bedin et al. (2008: 0.17/12.97) based on HST/ACS
photometry. When the new values (0.10/13.02) of NGC6791 from An et
al. (2009) are adopted, the colors and absolute magnitudes change
significantly. The coefficients in the calibrations are presented in
Table 4. It shows that dependence of absolute magnitude on
metallicity in the calibrations are increased significantly. For
example, metallicity dependence in the $\MV$ calibration is as high
as 0.41, which is not previously found in any calibration for RR
Lyrae stars and RC stars. For this reason, we do not adopt the
results based on new reddening in Table 4, which is presented in
this paper for reference.

\begin{table}
\caption{\bf The coefficients, $a$, $b$, $c$, and the scatters of
the calibrations in the formula of $M=a \feh + b t + c$. The
reddening and distance modulus of M71 are taken from Grundahl et
al. (2002) and the reddening from An et al. (2009) is used for
NGC6791.} \label{tb:sel} \setlength{\tabcolsep}{0.1cm}
\begin{tabular}{rrrrrrr}
\noalign{\smallskip}
\hline
filters  & $a$ & $b$ & $c$ & $\sigma_a$ & $\sigma_b$ & $\sigma$\\
\noalign{\smallskip}
$\Mu$ &     1.361 &  $-$0.055 &   4.428 &   0.109 &   0.063 &   0.151\\
$\Mg$ &    0.590 &  $-$-0.026&   1.807 &   0.055 &   0.032 &   0.076\\
$\Mr$ &    0.295 &   0.037 &   0.394 &   0.052 &   0.030 &   0.072\\
$\Mi$ &    0.148 &   0.055 &  $-$0.074 &   0.026 &   0.015 &   0.036\\
$\Mz$ &    0.112 &   0.088 &  $-$0.524 &   0.063 &   0.037 &   0.088\\
$\MV$ &    0.417 &   0.011 &   0.967 &   0.053 &   0.031 &   0.073\\
$\MR$ &    0.255 &   0.042 &   0.106 &   0.039 &   0.023 &   0.054\\
$\MI$ &    0.098 &   0.061 &  $-$0.601 &   0.037 &   0.022 &   0.052\\
\hline \noalign{\smallskip}
\end{tabular}
\end{table}

A few interesting results can be drawn from Table 3. First, it shows
that $\Mi$ has the smallest dependence on metallicity and age as
well as the smallest scatter. Second, the relations for $VRI$
filters are quite similar to those for $gri$ filters, but they show
less dependence on metallicity. Finally, the most interesting result
is that dependence of absolute magnitude on metallicity and age
shows different directions. The metallicity coefficient decreases,
while the age coefficient increases, as the filter varies from blue
$u$ to red $z$ in the formula of $M=a \feh + b t + c$. The $b$
coefficient increases when coefficient $a$ decreases; they show a
nonlinear relation  as shown in Figure 6. Meanwhile, the constant
$c$ decreases with decreasing $a$ coefficient, and they show a
linear trend.

\begin{figure}
\epsscale{1.0} \plotone{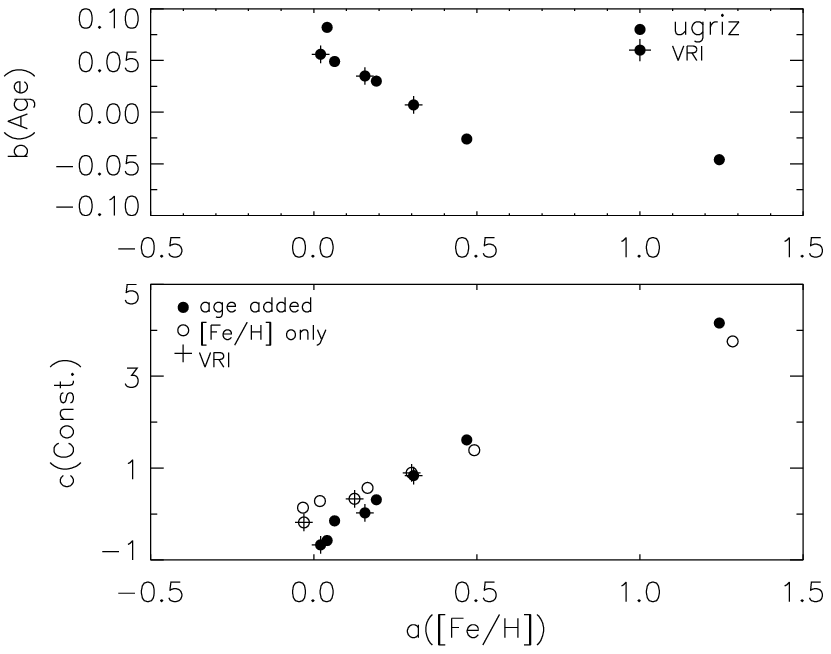} \caption{Trends of age coefficient
$b$ and constant $c$  vs. the metallicity coefficient $a$ in the
formula of $M=a \feh + b t + c$ (filled circles) and $M=a \feh + c$
(open circles). The additional plus indicates data based on $RVI$
filters.} \label{fig6}
\end{figure}

\subsection{Traditional calibrations with metallicity}
In order to investigate the age effect on the above calibrations,
we carry out the regressions of absolute magnitude with
metallicity in the traditional formula of $M=a \feh +c$.
The coefficients $a$ and the constants $c$ obtained
from the linear fittings to the data are presented in Table 5.
The results based on nonlinear calibrations in the formula of $M=a \feh +b
\feh ^2 +c$ are given in Table 6.

\begin{table}
\caption{The coefficients, errors  and scatters in the $M=a \feh +c$ formula.}
\label{tb:sel} \setlength{\tabcolsep}{0.1cm}
\begin{tabular}{rrrrr}
\noalign{\smallskip}
\hline
Filters  & $a$ &  $c$ & $\sigma_a$ & $\sigma$\\
\noalign{\smallskip}
$\Mu$ &    1.284 &   3.759 &   0.079 &   0.139 \\
$\Mg$ &    0.492 &   1.390 &   0.040  &  0.070\\
$\Mr$ &    0.165 &   0.569 &   0.038 &   0.066\\
$\Mi$ &    0.019 &   0.283 &   0.038 &   0.068\\
$\Mz$ &   $-$0.033 &   0.139 &   0.068 &   0.120\\
$\MV$ &    0.300 &   0.897 &   0.034 &   0.060\\
$\MR$ &    0.125 &   0.332 &   0.033 &   0.058\\
$\MI$ &   $-$0.030 &  $-$0.183 &   0.048 &   0.085\\
\hline \noalign{\smallskip}
\end{tabular}
\end{table}

\begin{table}
\caption{The coefficients, errors  and scatters in the
formula of $M=a \feh +b \feh ^2 +c$.}
\label{tb:sel} \setlength{\tabcolsep}{0.1cm}
\begin{tabular}{rrrrrrr}
\noalign{\smallskip}
\hline
Filters  & $a$ & $b$ & $c$ & $\sigma_a$ & $\sigma_b$ & $\sigma$\\
\noalign{\smallskip}
$\Mu$  & 1.688   & 0.313   & 3.667   & 0.144   & 0.103   & 0.087 \\
$\Mg$    & 0.507   & 0.011   & 1.387   & 0.115   & 0.083   & 0.070\\
$\Mr$    & 0.178   & 0.010   & 0.566   & 0.109   & 0.078   & 0.066\\
$\Mi$    & 0.034   & 0.012   & 0.279   & 0.111   & 0.080   & 0.068\\
$\Mz$   & 0.014   & 0.037   & 0.129   & 0.197   & 0.142   & 0.120\\
$\MV$    & 0.314   & 0.011   & 0.894   & 0.098   & 0.071   & 0.060\\
$\MR$    & 0.138   & 0.010   & 0.329   & 0.095   & 0.068   & 0.058\\
$\MI$    &$-$0.014   & 0.013   &$-$0.187   & 0.139   & 0.100   & 0.085\\
\hline \noalign{\smallskip}
\end{tabular}
\end{table}

It shows that $\Mu$ calibrations are slightly improved with smaller
scatters when the formula includes the metallicity only, and the
nonlinear calibration with metallicity is the best to reduce the
scatter from 0.14 to 0.09 mag. For the other four $griz$ filters,
nonlinear calibrations with metallicity are not necessary because
they give exactly the same scatters and the same constant $c$. As
compared with the calibrations with the age term included in Sect.
3.2, it shows that age plays a minor role to reduce the scatters and
the metallicity coefficient $a$ does not change too much while the
constant $c$ deviates from 0.2 to 0.7 mag as shown in Figure6.

In the comparison of the calibrations between narrow band $gri$
and wide band $VRI$ filters, it shows that the metallicity
coefficient reduces as the wide band is adopted, which does not
indicate less dependence on metallicity. As shown in the case of
$\Mi$ and $\MI$ calibrations in the the traditional formulae of
$M=a \feh +c$, the $a$ coefficient reduces from $0.02$ in the
$\Mi$ calibration to $-0.03$ in the $\MI$ calibration, but the
dependence on metallicity in the $\Mi$ calibration is smaller than
that in the $\MI$ calibration. When the age effect is included in
the calibrations, it shows that the metallicity dependence in the
wide band $VRI$ calibrations is smaller than that in narrow band
$gri$ filters. But the age coefficient and the constant in these
calibrations change accordingly. In fact, both the metallicity and
age dependence have a minimum at a particular wavelength band. The
minimum metallicity dependence is found at a band redder than $I$
and the minimum age dependence lies between $g$ and $r$ in the
$M=a \feh + b t + c$ calibrations.

\subsection{The comparison of absolute magnitudes from different stars}
$\Mg$ values of RHB at the metal poor edge are about 0.6-0.7 mag for
an old population with an age larger than 10 Gyr. These values are
quite consistent with theoretical $\Mg$ values of BHB as shown in
Figure 8 of Sirko et al. (2004) where $\Mg$ values of BHB are
0.6-0.7 mag for $(u-g)_0=1.1-1.2\,mag$. Note that $\umg$ values of
our RHB are also in the same range of 1.1-1.2 mag for the metal poor
and old clusters, while $\gmr$ values are about 0.3-0.4 mag for RHB
versus $-0.2$ mag for BHB at $\umg=1.1-1.2\,mag$. However, we notice
that $\Mg$ of RHB at $\gmr=0.3$ mag should be slightly fainter than
that of BHB at $\gmr=-0.2$ by 0.2 mag as we inspect the $\gmr$-$g_0$
diagrams of M5 (in Figure2),  M3 and NGC 7006 (in Fig 3). Moreover,
we notice that $g_0$ of BHB in M5 decreases by 0.4 mag as the $\gmr$
values vary from 0.0 to -0.2 mag and by 1.5 mag as the $\gmr$ values
vary from 0.0 to -0.5 mag , which is the color range that Sirko et
al. (2004) select BHB stars. But, the average $\Mg$ of BHB give the
same values of our RHB. This may explain the consistent $\Mg$ values
of our RHB with theoretical values of BHB presented in Sirko et al.
(2004). However, it is clear that $\Mg$ of BHB do vary with $\gmr$
colors. In this sense, absolute magnitudes of BHB stars should be
determined by taking into account their color variations.

In comparison with the calibration of $\MV=0.30 \feh +0.92$ by
McNamara (2001) based on RR Lyrae stars in clusters, deviations of
$\MV$(RHB-HB) are about 0.02 mag for clusters in group 1. To some
extent, these agreements may indicate that both $\Mg$ of RHB stars
derived in the present work and the translation relations of Chonis
\&  Gaskell (2008) are quite reliable. The deviations slightly
increase to 0.07-0.15 mag for clusters in group 2 due to their young
ages. Since the calibration by McNamara (2001) does not include the
age term and it is based on an old population, it may not be valid
for young populations. In this sense, new calibrations should be
established for metallicity higher than $\feh \sim -1.5$, where more
stars are younger in the context of Galactic evolution. In other
words, the age term is important in establishing these calibrations
in the metallicity range of $-1.5 <\feh< -0.5$.

For a metal rich population, RC stars are good distance indicators.
Recent work by Groenewegen (2008) gives the calibration of
$\MI=0.08(\feh+0.15)-0.26$ based on the revised Hipparcos parallaxes
by van Leeuwen (2007). The deviation of $\MI$ (RHB-RC) is 0.03 mag
for NGC6791 at $\feh=+0.4$ and 0.07 mag for M71 at $\feh=-0.7$.
Extending the relation of Groenewegen (2008) to low metallicity of
$\feh=-1.5$, the deviations of $\MI$ (RHB-RC) increase from 0.08 to
0.28 mag when age increases from 8.3 to 11.3 Gyr. The deviations of
$\MV$(RHB-HB) and $\MI$(RHB-RC) are shown in Figure 7, where
clusters in group 1 (open) and group 2 (filled) are shown
separately. Again, the deviations are mainly introduced due to age
differences between RC stars and globular clusters because $\MI$
calibrations are quite sensitive to age. Without the age term, $\MI$
calibrations cannot be established well as shown in Section 3.3,
where the uncertainty of metallicity coefficient and the scatter in
the $\MI$ calibration become quite large. It seems that absolute
magnitudes of BHB, RHB and RC are consistent when both metallicity
and age are taken into account.

\begin{figure*}
\epsscale{1.0} \plotone{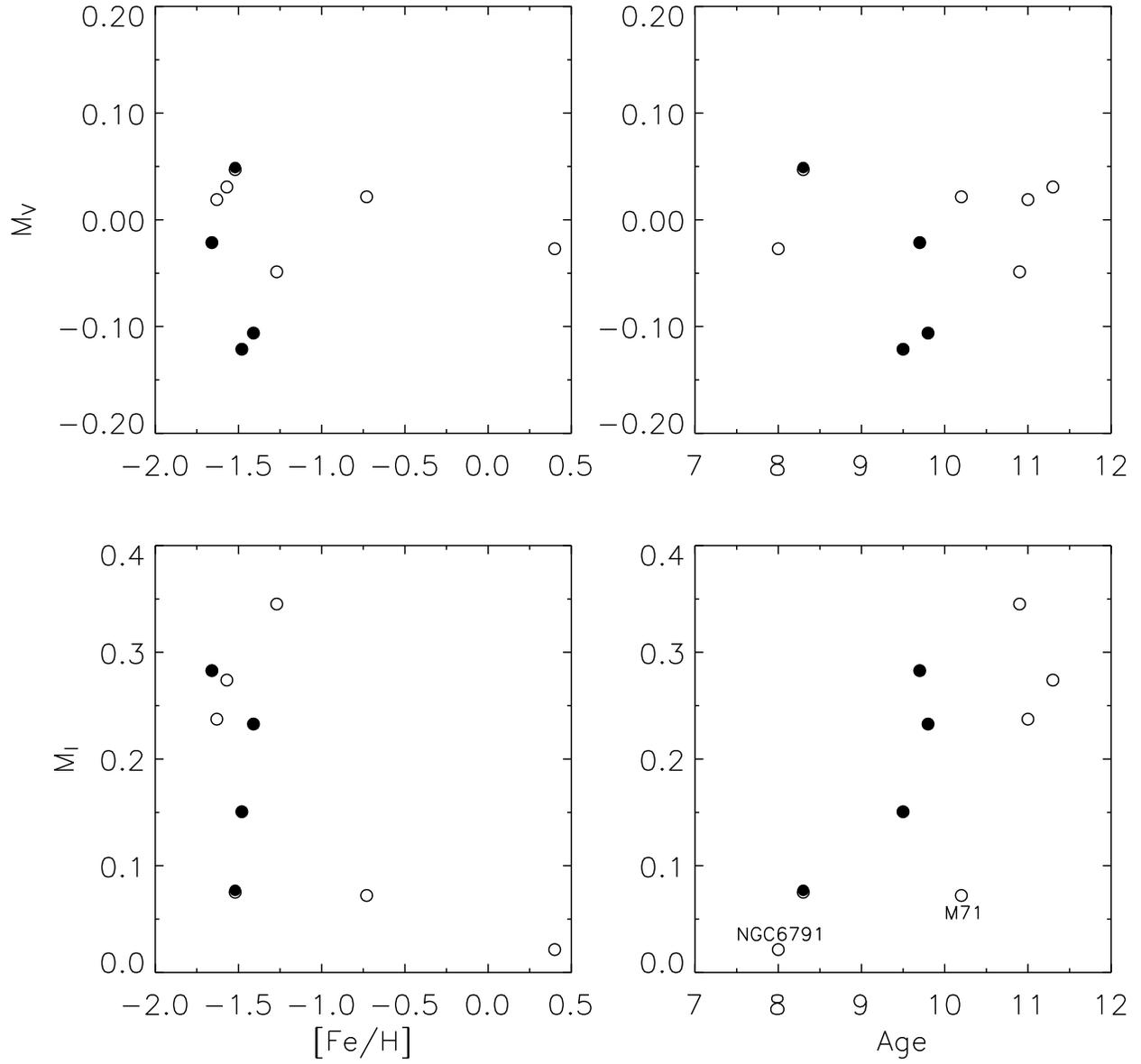} \caption{$\MV$ and $\MI$ of RHB stars
for clusters in group 1 (filled) and 2 (open) are compared with
those of BHB and RC stars in the literature.} \label{fig7}
\end{figure*}

\section{Implications and Conclusions}
\label{sect:Discussion}

Using the \gmr-$g_0$ diagrams of stars in the central parts of
clusters based on the SDSS survey from An et al. (2008), we have
selected RHB stars in nine clusters covering the metallicity of
$\-1.7 < \feh < +0.4$ and the age of 8-12 Gyr. Absolute magnitudes
of RHB stars in the SDSS $ugriz$ system are derived and they are
translated into the Johnson $VRI$ system based on the relations of
Chonis \&  Gaskell (2008). The calibrations of $\Mx$ (where
$X=u,g,r,i,z, V,R,I$) with metallicity and age are established and
they can be used to study the thick disk population which has the
same metallicity and age ranges as these clusters.

Based on an analysis of these data, we have found that $\Mi=
0.06\feh+0.040t+0.03$ is the calibration for distance determination
due to its smallest scatter of 0.04 mag when both metallicity and
age are known. On the other hand, it is also the best filter for
distance determination when both metallicity and age are unknown
because it has the least sensitivity to metallicity and age. The
scatters in $\Mg$ and $\Mr$ calibrations are also small (0.06 mag)
but their sensitivities to metallicity become larger. They can be
used for distance determination if the metallicity is known and the
metallicity range of the population is narrow. For a population with
known metallicity but large age span, the calibration in the $V$
filter may be better due to its smaller dependence on age. For a
population with both large metallicity and large span in age, e.g.
RC stars in the thin disk, the widely adopted $\MI$ filter may not
be the best choice. The deviations could be as large as 0.5 mag in
$\MI$ versus 0.25 mag in $\MV$ when metallicity varies from -0.5 to
0.0 and age varies from 0 to 8 Gyr, which is typical for stars in
the thin disk.

In particular, we have found that  coefficients of metallicity and
age in our calibrations show the opposite trends. This provides a
challenge to the previous suggestion that the calibration of
absolute magnitude in $I$ filter is better than that in the $V$
filter. Without the age term included in the calibration, it is
difficult to decide which is the best one. In the present work,
$\Mg$ and $\Mz$ show significant and opposite dependence on age. For
$\feh =-1.5$, an age difference of 3 Gyr introduces deviations of
0.2-0.4 mag in $\Mg$ and $\Mz$. Also, we provide the evidence on age
dependence of the $\MI$ calibration, which should be investigated in
the future.

\acknowledgments

We thank the anonymous referee for his valuable suggestions towards
improving the paper. Dr. James Wicker is thanked for careful
inspection on the written English of this paper. This study is
supported by the National Natural Science Foundation of China under
grants 10821061, 10673015, the National Basic Research Program of
China (973 program) No. 2007CB815103/815403, and by the Ministry of
Science and technology of China under grant 2006AA01A120.


\begin{thebibliography}{}
  \bibitem[An et al. (2008)]{An08} An, D., et al. 2008, ApJS, 179, 326
  \bibitem[An et al. (2009)]{An09} An, D., et al. 2009, ApJ, 700, 723
  \bibitem[Anthony-Twarog et al. (2007)]{Anthony07} Anthony-Twarog, B.J., Twarog, B.A., \& Mayer, L. 2007, 133, 1535
  \bibitem[Bedin et al. (2008)]{Bedin08} Bedin, L.R., et al. 2008, ApJ, 678, 1279
  \bibitem[Carney et al. (2005)]{Carney05}Carney, B.W., Lee, J.W. \& Dodson, B. 2005, AJ 129, 656
  \bibitem[Carretta et al. (2000)]{Carretta00} Carretta, E., Gratton, R.G., Clementini, G., Fusi Pecci, F. 2000, ApJ, 533, 215
 \bibitem[Carretta \& Gratton (1997)]{Carretta97} Carretta, E., \& Gratton, R. 1997, A\&AS, 121, 95
  \bibitem[Chonis \&  Gaskell(2008)]{Chonis08} Chonis, T.S. \&, Gaskell, C.M., 2008, AJ, 135, 264
   \bibitem[Feast 1997]{Feast97} Feast, M.F 1997, MNRAS, 284, 76
  \bibitem[Groenewegen (2008)]{Groenewegen08} Groenewegen, M.A.T. 2008, A\&A, 488, 935
  \bibitem[Grundahl et al. (2002)]{Grundahl02} Grundahl, F., Stetson, P.B., \& Andersen, M.I. 2002, A\&A, 395, 481
  \bibitem[Harris (1996)]{Harris96} Harris, W.E. 1996, AJ, 112, 1487
  \bibitem[Kraft \& Ivans (2003)]{Kraft03} Kraft R.P. \& Ivans I.I., 2003, PASP, 115, 143
  \bibitem[McNamara (2001)]{McNamara01} McNamara, D.H. 2001, PASP, 113,335
  \bibitem[Salaris \& Weiss (2002)]{Salaris02} Salaris, M \& Weiss, W. 2002, A\&A, 388, 492
    \bibitem[Sandage \& Tammann (2006)]{Sandage06} Sandage, A. \& Tammann, G.A. 2006, ARA\&A, 44, 93
  \bibitem[Sirko et al. (2004)]{Sirko04} Sirko, E., et al.  2004, AJ, 127, 899
  \bibitem[Stanek \& Garnavich (1998)]{Stanek98} Stanek, K.Z. \& Garnavich, P.M. 1998, ApJ, 503, L131
  \bibitem[van Leeuwen (2007)]{van07} van Leeuwen, F. 2007, A\&A, 474, 653
  \bibitem[van den Bergh (2006)]{van06} van den Bergh, S. 2006, Astron. J. 131, 1559
  \bibitem[Zhao et al. (2001)]{Zhao01}Zhao, G., Qiu, H., Mao, S.D. 2001, ApJ, 551, L85
  \bibitem[Zinn \& West (1984)]{Zinn84}Zinn, R., \& West, M.J. 1984, ApJS, 55, 45
\end{thebibliography}
\end{document}